\documentclass[twocolumn,showpacs,preprintnumbers,amsmath]{revtex4}

\usepackage{graphicx}
\usepackage{dcolumn}
\usepackage{bm}
\usepackage{amsmath}

\begin{document}

\preprint{APS/123-QED}

\title{Dewetting of thin polymer films near the glass transition}

\author{F. Saulnier} \email{florent.saulnier@college-de-france.fr}
\author{E. Rapha\"{e}l} \email{elie.raphael@college-de-france.fr}
\author{P.-G. de Gennes} \email{pgg@espci.fr}

\affiliation{%
Laboratoire de Physique de la Mati\`ere Condens\'ee, CNRS UMR 7125,
Coll\`ege de France\\
11, place Marcelin Berthelot, 75231 Paris Cedex 05, France.}

\date{28th of january, 2002}

\begin{abstract}
Dewetting of ultra-thin polymer films near the glass transition exhibits 
unexpected front morphologies [{\sc G. Reiter}, {\it Phys. Rev. Lett.}, {\bf 87}, 186101 
(2001)]. We present here the first theoretical attempt to understand these
features, focusing on the shear-thinning behaviour of these films. We analyse the
profile of the dewetting film, and characterize the time evolution of the dry region
radius, $R_{d}(t)$, and of the rim height, $h_{m}(t)$. After a transient time depending 
on the initial thickness, $h_{m}(t)$ grows like $\sqrt{t}$ while $R_{d}(t)$
increases like $\exp{(\sqrt{t})}$. Different regimes of 
growth are expected, depending on the initial film thickness and 
experimental time range. 

\end{abstract}

\pacs{68.60.-p, 68.15.+e, 68.55.-a, 83.10.-y}

\maketitle
     
Thin liquids films are ubiquitous entities in science and technology.
In engineering, for instance, they serve to protect surfaces,
and applications arise in paints, adhesives and membranes \cite{oron}.
Thin liquids films display a variety of interesting dynamics phenomena
and have therefore been the focus of many experimental and
theoretical studies \cite{revues}.
Nearly half a century ago, Taylor \cite{taylor} and Culick \cite{culick} analyzed
the growth of a circular hole in a thin liquid sheet \cite{dupre}.
By balancing surface tension
forces against inertia, they found that the rim of liquid at the edge
of the films retracts at a constant velocity, a prediction first checked
experimentally by Mc Entee and Mysels \cite{mysels}.
The precise shape of the rim was later analysed by Keller {\it et al.} \cite{keller,koplik} who
showed that it is a cylindrical cap expanding in time like $\sqrt{t}$.
Recent experiments by Debr\'egeas and collaborators \cite{debregeas}
on thin suspended films of very viscous liquids revealed unexpected features:
(a) First, the retraction velocity grows exponentially
with time (with a characteristic time scale $\tau_{i} = h_{i} \eta /|S|$, 
where $h_i$, $\eta$ and $S$ are respectively the initial film thickness, the viscosity
and the spreading coefficient \cite{rem1}), (b) second, the
liquid is {\it not} collected into a rim and the film remains flat through
the retraction. According to these authors, the uniform thickening of
the retracting film was a consequence of its viscoelasticity, which permits an elastic
propagation into the film of the surface tension forces acting on the edge.
Solving numerically the Navier-Stokes
equations for long wavelength modulations of the film, Brenner and Gueyffier \cite{brenner} 
showed, however, that the absence of rim can also 
result from a purely viscous effect.

\begin{figure}
\resizebox{0.40 \textwidth}{!}{%
  \includegraphics*[3.4cm,11cm][18cm,18.2cm]{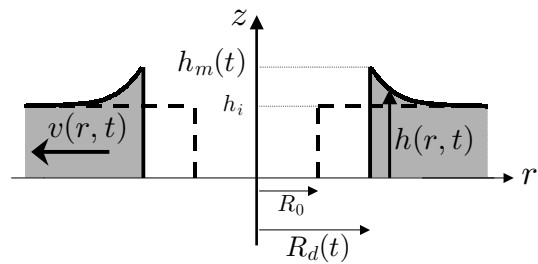} }
\caption{Film geometry: $h(r,t)$ is the profile of the film, $h_{m}(t)$ is the height of the 
rim, and $R_{d}(t)$ is the radius of the dry
zone. The initial ($t = 0$) step-like profile
is represented by the dashed line. 
$v(r,t)$ is the radial, axisymmetric flow field \cite{rem2}.} 
\label{dessin modele}
\end{figure}

Very recently \cite{reiter}, Reiter studied the dewetting of ultrathin
({\it i.e.} thinner than the coil size), almost glassy
polystyrene (PS) films deposited onto silicon wafers coated with a polydimethylsiloxane 
(PDMS) monolayer. He found that a highly asymmetric rim, with an extremely steep side towards the
interior of the hole and a much slower decay on the rear side,
builds up progressively \cite{herminghaus}. In this Letter we present the first theoretical attempt to understand these
new features. Figure \ref{dessin modele} shows the film geometry.

In order to characterise the rheologic properties of the film, we introduce the stress
tensor $\sigma_{ij}=-p 
\delta_{ij}+\sigma_{\: \: \: ij}^{m}$, where $p$ is the pressure and $\sigma_{\: \: \: ij}^{m}$ represents the 
effects of internal friction.
We relate $\sigma_{\: \: \: ij}^{m}$ to the strain rate tensor
$\dot{\gamma}_{ij}$ by a constitutive law of the form:

\begin{equation}
\sigma^{m} (\dot{\gamma}) = \sigma_{0} \: \Phi (\dot{\gamma} \tau) 
\label{rheolaw}
\end{equation}

\begin{figure}
\resizebox{0.48 \textwidth}{!}{%
  \includegraphics*[1.8cm,8.5cm][19cm,22.5cm]{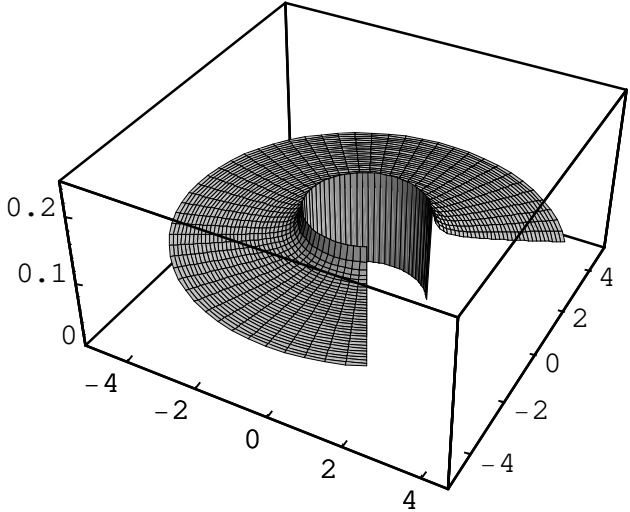} } \caption{Film profile
for $h_{i}=0.1 h^{*}$ and $t=10^{-2} \tau$. For
clarity reasons, a sectional view is displayed on a quarter of
circle.} \label{profil3d}
\end{figure}        
            
where $\sigma_{0}$ and $\tau$ are material constants, and $\Phi$ is a generic function.
For a purely viscous liquid, the function $\Phi$ is linear, whereas for polymers,
just above $T_{g}$, it can be shown within the
framework of the free-volume model
\cite{aparaitre} that $\sigma^{m}$ is expected to vary logarithmically with ${\dot{\gamma}}$
as \cite{dyre}:

\begin{equation}
\Phi(\dot{\gamma} \tau) = \ln(1+\dot{\gamma} \tau)
\label{shear-thinning}
\end{equation}

At low strain-rates
($\dot{\gamma} \tau <1$), this law
displays a viscous-type behaviour ($\sigma^{m}
\approx \eta_{0} \dot{\gamma}$, with a zero-shear viscosity $\eta_{0}=\sigma_{0} \tau$),
while for large values of $\dot{\gamma} \tau$, $\sigma^{m}$ reaches an almost constant value
(shear-thinning behaviour).            

Assuming the fluid to be incompressible, mass conservation leads to:

\begin{equation}
\frac{1}{h(r,t)} (\frac{\partial h(r,t)}{\partial
t}+\frac{r \beta(r,t)}{\tau} \frac{\partial h(r,t)}{\partial
r})=\frac{\alpha(r,t)-\beta(r,t)}{\tau} \label{eqn hrt}
\end{equation}
                                      
Equation \ref{eqn hrt} involves two unknown functions $\alpha(r,t)$ and 
$\beta(r,t)$, that are positive, dimensionless forms of the strain-rate components
$\dot{\gamma}_{rr}$ and $\dot{\gamma}_{\phi \phi}$:

\begin{equation}
\left\{ \begin{array}{ll} \alpha = -\tau\dot{\gamma}_{rr}=-\tau
\frac{\partial
v}{\partial r} \\
\beta = \tau \dot{\gamma}_{\phi \phi}=\tau \frac{v}{r}
                    \end{array}
            \right.
\label{definition}
\end{equation}

We thus need two additional equations to determine $h(r,t)$. First, note that 
the following partial differential equation can be directly derived
from Eq.\ref{definition}:

\begin{equation}
\frac{\partial \beta}{\partial r}= - \frac{\beta +\alpha}{r} 
\label{eqn suppl}
\end{equation}
                      
Neglecting the inertial term, conservation of momentum
(projected on the radial direction \cite{christensen}) leads to:

\begin{equation}
\frac{\partial \sigma_{rr}}{\partial r}
+\frac{\sigma_{rr}-\sigma_{\phi \phi}}{r}=0
\label{motion}
\end{equation}                                           
                  
It can be shown that Eqs. \ref{eqn suppl} and \ref{motion}, along with the free-surface
boundary condition ({\it i.e.} $\sigma_{zz}=0$ at the
contact with ambient atmosphere \cite{rem4}), allow one to express 
the strain rate $\alpha$ as a function of $\beta$ only : 
$\alpha=F(\beta)$. Substitution of the constitutive law \ref{rheolaw}
in Eq.\ref{motion} then leads to the following differential equation for $F$:

\begin{equation}
 \frac{\mbox{d}[\Phi(F(\beta))-\Phi(F(\beta)-\beta)]}{\mbox{d} \beta} =
\frac{\Phi(F(\beta))-\Phi(\beta)}{F(\beta) + \beta}
\label{f(beta)}
\end{equation}
                                
which has to be solved along with the condition $F(0)=0$
(far away from the perturbed central region, the strain rates must decrease to 
zero).

In order to solve Eq.\ref{eqn 
hrt}, we should supply it with initial and boundary conditions.
Our initial profile is assumed to be uniform, with a thickness $h_i$, except in a bored
region ranging from $r=0$ to $r=R_{0}$ (as shown in
fig.\ref{dessin modele}). $R_{0}$ is our characteristic radial
length used thereafter to make $r$ dimensionless: 

\begin{equation}
h(r,t=0)= \left\{ \begin{array}{ll} h_{i} & \mbox{if $r \geq R_{0}=1$} \\
                                    0 & \mbox{otherwise}
                    \end{array}
            \right.
\end{equation}

\begin{figure}
\resizebox{0.47 \textwidth}{!}{%
  \includegraphics*[1.6cm,10.5cm][19.5cm,21.4cm]{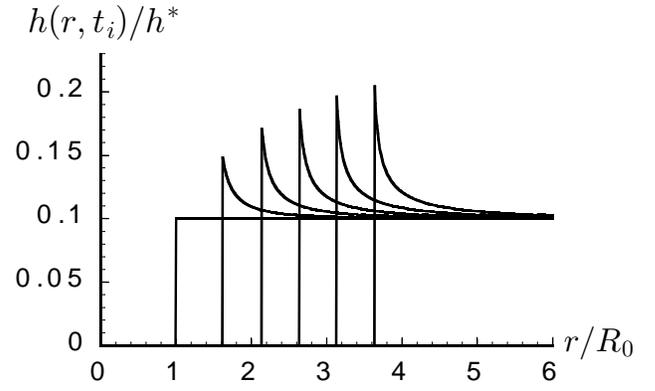} }
\caption{Evolution of the film profile $h(r,t)$ for $h_{i}=0.1 h^{*}$ and
$t$ ranging from $t=0$ to $t=5.10^{-2} \tau$ with a time step of $10^{-2} \tau$.} 
\label{profil qui monte}
\end{figure}             

We do not consider here the
origin of the initial dewetting process: experimentally, it is found that a
thick PS film on a silicon substrate
is metastable and dewets via nucleation and growth of dry patches
\cite{redon}, while thinner films ($h_{i}<$100 nm) are unstable and dewet by
spinodal decomposition \cite{gunterprec}.
In his latest experiments on ultra-thin films \cite{reiter},
Reiter characterized the early
stage of the dewetting process by the formation and coalescence of
little holes, with the displaced material uniformly distributed
between the holes, without visible rims. Our initial time $t=0$
might correspond to the end of this preliminary process.
                                  
Equation \ref{eqn hrt} applies outside the dewetted region ($R_{d}(t)  \leq r< \infty$).
At the edge of the rim ($r=R_{d}(t)$), the rim height, $h_{m}(t)$, 
can be determined by taking into account capillary forces.         
The applied force on the rim, pushing the film away the dry area, must be
balanced by the internal radial stress: $|S|=|\sigma_{rr}| h_{m}(t)$.
Assuming the lateral extension of the film to be large enough,
the film thickness must
reach $h_{i}$ far from the dry region:
$\lim_{r \rightarrow \infty}h(r,t)=h_{i} \; \;   (\forall t)$. 
The complete resolution of our set of equations
\ref{eqn hrt}-\ref{eqn suppl}-\ref{f(beta)}
can be achieved using a method of characteristics \cite{aparaitre}.
Thereafter, all thicknesses will be made dimensionless by normalizing with a characteristic length
$h^{*} \equiv |S|/\sigma_{0}$
(for Reiter's experimental conditions, we estimate $h^{*}$ to be of the order of 500 \AA).

Note first that for a purely viscous liquid ($\Phi (\dot{\gamma} \tau)=\dot{\gamma} \tau$),
our model leads to a constant and uniform thickness
for the film, with an exponential growth of the dry radius: 

\begin{equation}
R_{d}(t) = R_{0} e^{\frac{|S|t}{\sigma_{0} \tau h_{i}} } = R_{0} e^{\frac{t}{\tau_{i}}}
\label{reg visqueux}            
\end{equation}                        
                        
This is in complete agreement with the experimental results of Debr\'egeas
{\it et al.} for the dewetting of 
suspended polymer films
\cite{debregeas} and supported films \cite{debregeasmacromol}.

If we take into account the shear-thinning behavior of the polymer 
film ($\sigma^{m} (\dot{\gamma}) = \sigma_{0} \ln(1+\dot{\gamma} 
\tau)$)
the dewetting front is quite different. An example of profile obtained for $h_{i}=0.1 h^{*}$ is shown in
Fig.\ref{profil3d}. This profile is 
characterized by a highly asymmetric shape for the rim.
There is a striking similarity
between such a profile and those observed by Reiter
for ultra-thin films \cite{reiter}. 
On fig.\ref{profil qui monte} we present the time evolution of this
profile (for $t$ ranging from $t=0$ to $t=5.10^{-2} \tau$, with a time step of $10^{-2} \tau$).
The velocity field $v(r,t)$ is found to be almost proportional to $1/r$ at any time for thick profiles ($h_{i}
\gg h^{*}$), similar to the long range radial plug
flow observed by Debr\'egeas {\it et al.} \cite{debregeas}. We now focus on two natural parameters
characterizing the film profile: the dry radius $R_{d}(t)$ and rim height $h_{m}(t)$.

\begin{figure}
\resizebox{0.45 \textwidth}{!}{%
  \includegraphics*[2cm,9.4cm][14.5cm,16.8cm]{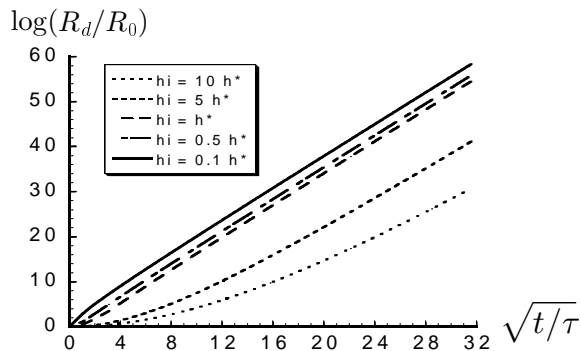} }
  \caption{Logarithmic plot
  of the dry zone radius $R_{d}$ versus $\sqrt{t/\tau}$, for $h_{i}=$ 0.1, 0.5, 1, 5 and 10 
  $h^{*}$. Note the linear behaviour of $\log{[R_{d}(t)/R_{0}]}$ with $\sqrt{t/\tau}$
  after a transient time.}
\label{rayon sec t05}
\end{figure} 

\begin{figure*}
\resizebox{0.72 \textwidth}{!}{%
  \includegraphics*[1cm,20.5cm][20cm,27.2cm]{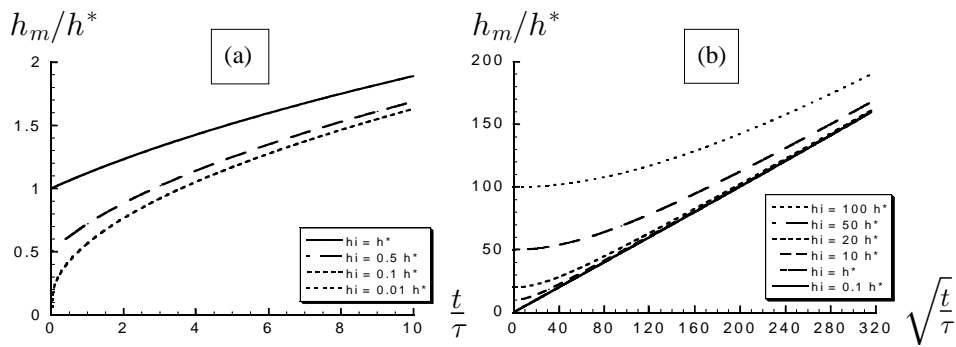} }
  \caption{Time evolution of the rim height $h_{m}$:
  {\bf (a)} Plot of $h_{m}$ versus $t/\tau$, for $h_{i}=$ 0.01, 0.1, 0.5 and 1 $h^{*}$. Note that the curves
  corresponding to $h_{i}=$0.1 and 0.01$h^{*}$ cannot be distinguished within the graph
  scale. {\bf (b)} Plot versus $\sqrt{t/\tau}$
  to show the linear behaviour of $h_{m}$ with $\sqrt{t/\tau}$ after the transient time $t_{0}$.
  For a thick film ($h_{i} \gg 
  h^{*}$), two regimes can be distinguished.
  For $t < t_{0} \sim \tau (\frac{h_{i}}{h^{*}})^{2}$,
  $h_{m}$ is constant and $R_{d}(t)$ increases exponentially with time.
  For $t> t_{0}$, $h_{m}$ growths linearly
  with $\sqrt{t/\tau}$ and $R_{d}(t)$ expands like $\exp{\sqrt{t/\tau}}$.}
  \label{hm en t^05}
\end{figure*}

In order to discuss the
time dependence of the rim height, we need to compare $h_{i}$ with the characteristic thickness
$h^{*}$ (in Fig. \ref{hm en t^05} we present our results for different
initial thicknesses). For very thin profiles ($h_{i} \ll h^{*}$), the increase of rim
height is initially very fast, whereas for thicker films ($h_{i}
\gg h^{*}$), $h_{m}(t)$ is nearly constant during a long period of
time, before growing faster. For the sake of conciseness, we will mainly focus on the case of thick films,
for which simple scaling laws can be analytically obtained. As shown in Fig.\ref{hm en 
t^05}, for $h_{i}=100 h^{*}$
two different time regimes can be distinguished: at the beginning of hole 
formation,
$h_{m}$ is nearly constant and equals its initial value $h_{i}$. 
We have checked numerically (cf. Fig.\ref{hm en t^05} {\bf b}) that the anticipated analytical 
behaviour

\begin{equation}
h_{m}(t) \underset{t \gg t_{0}}{\sim} \frac{h^{*}}{2}
\sqrt{t/\tau} \label{hm gd tps}
\end{equation}

is obeyed after a crossover time $t_{0} \approx \tau
(\frac{h_{i}}{h^{*}})^{2}$

In Fig.\ref{rayon sec t05}, we present the time evolution of $R_{d}$
for different initial thicknesses $h_{i}$. It is
physically understandable that $R_{d}$, at a given time $t$, is larger
for thinner films : for a given applied force $|S|$ per unit
length, the thicker the film, the more the material to be
displaced and the lower the dewetting velocity. Fig.\ref{rayon sec t05}
shows that the following anticipated analytical behaviour is obeyed for $t \gg t_{0}$:
                   
\begin{equation}
R_{d}(t) \underset{t \gg t_{0}}{\sim} R_{\infty} e^{4
\sqrt{t/\tau}} 
\label{rayon sec a t g tau}
\end{equation}  

Note that the coefficient $R_{\infty}$
strongly depends on $h_{i}$. During the early stage ($t \ll t_{0}$), the strain rates
at the rim are constant and small, as a consequence of
the constant large thickness. In this case,
the rheological law of the film is viscous-type and the dry radius increases exponentially, as 
for the case previously discussed (see Eq.\ref{reg visqueux}).
The range of thicknesses experimentally studied by Debr\'egeas {\it et al.} (5 to 250 $\mu$m) covers the
domain $h_{i} \gg h^{*}$. As the corresponding crossover time is
very large ($t_{0} \gg \tau$), these experiments only covered mainly the first regime $t \ll t_{0}$
(this criterion $h_{i} \gg h^{*}$ is in fact a basic hypothesis of their ``soft balloon'' model
\cite{rem5}). The crossover to the exp($\sqrt{t}$) regime is a consequence of
the non-linearity of rheological law \ref{shear-thinning}.

The case $h_{i} \ll h^{*}$ will not be fully discussed here (cf. \cite{aparaitre}).
Briefly, we can note on Fig.\ref{hm en t^05} that a sharp rim grows very quickly
(with an initial speed $\dot{h}_{m} \approx h_{i} \exp{h^{*}/2
h_{i}}$) at short times. In this regime, $h_{m}(t)$ is well described 
by the integral equation:

\begin{equation}
\int_{\frac{h^{*}}{2 h_{m}(t)}}^{\frac{h^{*}}{2 h_{i}}} \frac{e^{-x}}{x}
\mbox{d}x \underset{t \ll \tau}{\approx} \frac{t}{\tau}
\end{equation}
                     
Later on, $h_{m}$ and $R_{d}$ reach the same laws of growth as in the case of thick films
(Eqs.\ref{hm gd tps} and \ref{rayon sec a t g tau}).    

In conclusion, our model accounts well for the exponential growth 
and the absence of rim characteristics of the dewetting regime observed by Debr\'egeas 
{\it et al.} for viscous polymer films. Taking into account the shear-thinning behaviour of 
the polymers near $T_{g}$ enables us to see the modifications 
induced by this particular rheology on the film morphology \cite{dalnoki}.
It appears from our results that the early stages 
of dewetting for a shear-thinning polymer film are mainly 
determined by the ratio of its initial thickness $h_{i}$ to a 
characteristic scale $h^{*}$ (related to the driving force of the process, $S$, and 
the rheological response of the material, characterized by $\sigma_{0}$). 
In all cases, after a transient time depending on $h_{i}$, the rim 
height grows proportionally to $\sqrt{t/\tau}$ while 
the dry radius expands like $\exp{(4\sqrt{t/\tau})}$. The film profile
exhibits a sharp, asymmetric rim similar to the one 
observed by Reiter in his latest experiments with ultrathin PS films.
Work is now in progress to incorporate in our model the Laplace pressure (see \cite{rem4}) which might
lead to oscillatory dewetting fronts \cite{brenner,herminghaus}, and to evaluate 
the possible effects of chain entanglements. In the future, we aim to study the
dewetting of thin polymer films {\it 
below} $T_{g}$, where the existence of a yield stress in the rheological
response of the material may lead to new dewetting morphologies.

We wish to acknowledge fruitful discussions with {\sc A. Aradian}, {\sc G. Debr\'egeas}, and {\sc G. Reiter}.


\begin{thebibliography}{9}
\bibitem{rem1}
The driving force of the dewetting process is a difference of interfacial
tensions: $-S = \gamma_{SL}+\gamma_{LO}-\gamma_{SO} >0$ (where SL
corresponds to the solid/liquid interface, SO to solid/air
interface, etc...).
\bibitem{rem4}
In order to keep the model analytically tractable, we here neglect the pressure
discontinuity arising from the interface curvature. Further work 
will be needed to reassess the range of validity of this 
assumption.
\bibitem{rem5}
It is clear that this exponential law cannot be valid for too thin films. Writing
$\eta = E \tau_{rep}$ (where $E$ is an elastic modulus, and
$\tau_{rep}$ is the reptation time of polymers, the relaxation
time of the material), one obtains $\tau_{i} =\tau_{rep} h/h_{0}$ with
$h_{0} \equiv |S|/E$, which leads to a characteristic time $\tau$
smaller than $\tau_{rep}$ if $h_{i} < h_{0}$: the relaxation would
be too fast.
\end{thebibliography}
\end{document}